\documentclass[prd,showpacs,amsmath,amssymb]{revtex4}
\usepackage{graphicx}
\usepackage{bm}
\usepackage{subfigure}
\usepackage{slashed}
\usepackage{mathrsfs}
\begin{document}
\title{\boldmath Isospin violating decay of $\psi(3770)\rightarrow J/\psi + \pi^0$\unboldmath}
\author{Ze-kun Guo$^1$\footnote {E-mail: guozk@ihep.ac.cn}, Stephan Narison$^2$\footnote {E-mail: snarison@yahoo.fr}, Jean-Marc Richard$^3$\footnote {E-mail: j-m.richard@ipnl.in2p3.fr}, and Qiang Zhao$^{1,4}$\footnote {E-mail: zhaoq@ihep.ac.cn}}
\affiliation{1) Institute of High Energy Physics, Chinese Academy of
Sciences, Beijing 100049, P.R. China \\
2) Laboratoire Particules et Univers de Montpellier, CNRS-IN2P3,
Case 070, Place Eug\`{e}ne Bataillon, 34095 - Montpellier Cedex 05, France \\
3) Universit\'{e} de Lyon et Institut de Physique
Nucl\'{e}aire de Lyon, IN2P3-CNRS-UCBL \\
4, rue Enrico Fermi, F-69622 Villeurbanne, France \\
4) Theoretical Physics Center for Science Facilities, CAS, Beijing
100049, P.R. China }
\date{\today}
\begin{abstract}
The strong-isospin violation in $\psi(3770)\rightarrow J/\psi +
\pi^0$ via intermediate $D$ meson loops is investigated in an
effective Lagrangian approach. In this process, there is only one
$D$-meson loop contributing to the absorptive part, and the
uncertainties due to the introduction of form factors can be
minimized. With the help of QCD spectral sum rules (QSSR), we
extract the $J/\psi DD^*$ form factor as an implement from the first
principle of QCD. The $DD^*\pi^0$ form factor can be well determined
from the experimental data for $D\rightarrow\pi l\nu$. The
exploration of the dispersion relation suggests the dominance of the
dispersive part via the intermediate $D$ meson loops even below the
open charm threshold. This investigation could provide further
insights into the puzzling question on the mechanisms for
$\psi(3770)\to$ non-$D\bar{D}$ transitions.
\end{abstract}
\maketitle
\section{Introduction}\label{sec:0}
The non-$D\bar{D}$ decays of $\psi(3770)$ have attracted a lot
attention during the past decades. As $\psi(3770)$ is the first
state above the open charm $D\bar{D}$ threshold, its decay was
believed to be saturated by the $D\bar{D}$ channel via the
Okubo-Zweig-Iizuka (OZI) connected diagram. Such an anticipation was
supported by early experimental data which showed that exclusive
decays of $\psi(3770)\to$ non-$D\bar{D}$ were negligibly small.
Theoretical calculations of the perturbative QCD (pQCD) leading
order contributions also suggested rather small non-$D\bar{D}$
branching ratios for
$\psi(3770)$~\cite{Kuang:1989ub,Ding:1991vu,Rosner:2001nm,Rosner:2004wy,Eichten:2007qx,Voloshin:2005sd}.

Interestingly, recent studies of the $\psi(3770)$ non-$D\bar{D}$
decays in experiment and theory have exposed unexpected results
which complicated the situation. In experiment, the $D\bar{D}$ cross
section measurement by the CLEO collaboration suggests that the
non-$D\bar{D}$ branching ratio is consistent with zero with an upper
limit of  about 6.8\% \cite{He:2005bs,:2007zt,Besson:2005hm}. Rather
contradicting the CLEO results, the BES collaboration finds much
larger non-$D\bar{D}$ branching ratios of $\sim 15\%$ in the direct
measurement of non-$D\bar{D}$ inclusive cross
section~\cite{bes-3770}. Recently a next-to-leading-order (NLO)
nonrelativistic QCD (NRQCD) calculation of the $c\bar{c}$
annihilation width for $\psi(3770)$ suggests that the higher order
contributions can account for about $5\%$ of the $\psi(3770)$
non-$D\bar{D}$ decay branching ratios at most~\cite{He:2008xb}. In
Refs.~\cite{Zhang:2009kr,Liu:2009dr}, it was proposed that the
open-charm threshold effects via intermediate meson loops (IML)
could serve as an important nonperturbative mechanism to produce
sizable non-$D\bar{D}$ branching ratios. Note that $\psi(3770)$ is
close to the $D\bar{D}$ open threshold. A natural conjecture is that
the $D\bar{D}$ threshold would play an important role in its
production and decay. This mechanism turns out to be successful in
the explanation of the decay of $\psi(3770)\to
\text{vector}+\text{pseudoscalar}$ as one of the non-$D\bar{D}$
decay channel of $\psi(3770)$~\cite{Zhang:2009kr,Liu:2009dr}.

During the past few years, there have been observations of a large
number of heavy quarkonium states~\cite{Olsen:2009ys} at the
B-factories (Belle and BaBar) and electron storage-rings (CLEO).
Some of those states have masses close to open thresholds and cannot
be easily accommodated in the framework of potential quark models.
For instance, the well-established $X(3872)$ is located in the
vicinity of $D^*\bar{D}$ threshold and its mass as a $1^{++}$ state
is nearly 100 MeV lower than the first radial excitation of
$\chi_{c1}$ in  potential models. Such observations, on the one
hand, have raised serious questions on the constituent degrees of
freedom within heavy quarkonia, and on the other hand, raised
questions on the role played by the open decay thresholds via the
IML as an important nonperturbative mechanism in the understanding
of the properties of those newly observed states. Such a mechanism
symbolizes a general dynamical feature in the charmonium mass
region, thus should be explored broadly in various processes.

To gain further insights into the underlying dynamics and understand
better the properties of the IML, we are motivated to study the
decays of $\psi(3770)\to J/\psi+\eta$ and $J/\psi+\pi^0$. First, we
note that the decay of $\psi(3770)\to J/\psi+\eta$ is one of few
measured non-$D\bar{D}$ decay channels in experiment with
$BR(\psi(3770)\to J/\psi+\eta) = (9\pm 4)\times
10^{-4}$~\cite{Nakamura:2010zzi}. One can estimate the branching
ratio of $\psi(3770)\to J/\psi+\pi^0$ via $\eta$-$\pi^0$ mixing
based on the leading-order chiral perturbation theory. The mixing
intensity can be expressed as~\cite{Gross:1979ur}
\begin{eqnarray}
  \epsilon_0=\frac{\sqrt{3}}{4}\frac{m_d-m_u}{m_s-(m_u+m_d)/2}.
\end{eqnarray}
Using Dashen's theorem~\cite{Dashen:1969eg}, one obtains
\begin{eqnarray}
\epsilon_0=\frac{1}{\sqrt{3}}\frac{M_{K^0}^2-M_{K^+}^2+M_{\pi^+}^2-M_{\pi^0}^2}{M_\eta^2-M_{\pi^0}^2}=0.01
\ .
\end{eqnarray}
Taking the $\eta-\eta^\prime$ mixing into account, the mixing
intensity is slightly enhanced~\cite{Kroll:2005sd}
\begin{eqnarray}
\hat{\epsilon}=\epsilon_0\sqrt{3}\cos\phi,
\end{eqnarray}
where $\sqrt{3}\cos\phi=1.34$ would be unity if $\phi$ is the ideal
mixing angle. With the $\eta$-$\pi^0$ mixing intensity in a range of
$0.01\sim 0.02$, the  branching ratio of $\psi(3770)\to
J/\psi+\pi^0$ from $\eta$-$\pi^0$ mixing is at most the order of
$10^{-6}$. This result actually sets up a limit for the
$\eta$-$\pi^0$ mixing contributions in $\psi(3770)\to J/\psi+\pi^0$.
In contrast,  the Particle Data Group
(PDG2010)~\cite{Nakamura:2010zzi} gives an experimental upper limit
$BR(\psi(3770)\to J/\psi+\pi^0)< 2.8\times 10^{-4}$. A recent
investigation of $\psi'\to
J/\psi+\pi^0$~\cite{Guo:2009wr,Guo:2010ak} based on a
nonrelativistic effective field theory (NREFT) suggests that the
strong-isospin violation via the IML is relatively enhanced by $1/v$
in comparison with the tree-level contribution where the pion is
emitted directly from the charmonium through soft gluon exchanges,
where $v\simeq 0.5$ is the velocity of the intermediate charmed
meson. Since  $\psi(3770)$ is close to the $D\bar{D}$ threshold, we
expect that such a strong-isospin mechanism would also play a role.
As a consequence, the IML mechanism may lead to a sizable branching
ratio of $\psi(3770)\to J/\psi+\pi^0$ which might be significantly
larger than that given by the $\eta$-$\pi^0$ mixing. In an early
study~\cite{Achasov:1990gt,Achasov:1994vh,Achasov:2005qb}, the
absorptive contribution from the intermediate $D\bar{D}$ in the
decay of $\psi(3770)\to J/\psi+\pi^0(\eta)$ was calculated with an
exponential form factor determined by the characteristic mass scale
in the exchange channel.  It was also argued that the real part
contribution from $D\bar{D}$ and other heavier $D$ meson loops
should cancel each other in order to obey the OZI rule successfully
in $J/\psi$ decay. However, it is found~\cite{Wang:2012mf} that the
IML effects may still be important in the decay of charmonium close
to the open charm threshold. This is because the quark-hadron
duality turns out to have been broken locally. As a consequence, the
decay of a charmonium state can still experience the open threshold
effects significantly if its mass is close to the open threshold.
Such a scenario may imply that the real parts of the exclusive
$\psi(3770)$ decays could not be neglected and could explain the
observed sizeable non-$D\bar{D}$ branching ratios of
$\psi(3770)$~\cite{bes-3770}.


In this work, we shall apply an effective Lagrangian approach (ELA)
to investigate the IML effects in $\psi(3770)\to J/\psi+\pi^0$, and
demonstrate that the IML transitions have dominant contributions to
this isospin-violating decay channel. As an important improvement of
this approach, we shall implement form factors from QCD spectral sum
rules (QSSR) for the off-shell $J/\psi D D^*$ coupling vertex, while
the $D^*D\pi^0$ form factor can be extracted from the semileptonic
decay of $D\to \pi^0 l\nu$. We mention in advance that this
elaborate treatment will allow a reliable estimate of the absorptive
amplitude of $\psi(3770)\to J/\psi+\pi^0$. Meanwhile, we also
explore the real part of the IML contributions with the help of
dispersion relation, within which the effective threshold is
determined by the experimental data of $\psi(3770)\to J/\psi+\eta$.

This paper is organized as follows: In Sec.\ \ref{sec:1} the ELA for
the IML transitions is formulated. In Sec.\ \ref{sec:2} the form
factors from QCDSR and $D$ meson semileptonic decays are
investigated. The numerical results for the isospin-violating decay
$\psi(3770)\to J/\psi+\pi^0$ are presented in Sec.\ \ref{sec:3}, and
a brief summary is given in Sec.\ \ref{sec:4}.

\boldmath\section{Isospin-violating decay of  $\psi(3770)\rightarrow
J/\psi+\pi^0$ via IML transitions }\unboldmath \label{sec:1}
\subsection{Absorptive part}\label{sec:1.1}
\begin{figure}[!htp]
\centering
\includegraphics[width=0.5\textwidth]{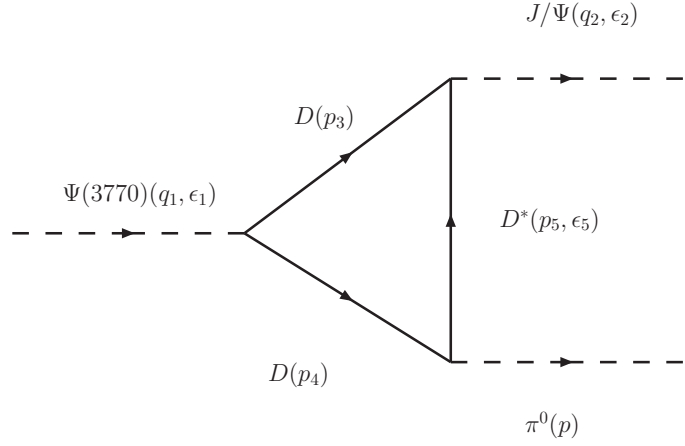}
\caption{The $D$-loop diagram contributing to the absorptive part.}
\label{fig:FeynmanDiagram}
\end{figure}

As illustrated by Fig.~\ref{fig:FeynmanDiagram}, only the $DD(D^*)$
loop (the meson in the parenthesis denotes the exchanged particle
between $J/\psi$ and $\pi^0$) contributes to the imaginary part. The
effective Lagrangians  are  as
follows~\cite{Colangelo:2003sa,Casalbuoni:1996pg},
\begin{eqnarray}
\mathcal{L}_{\psi'' DD}&=&-ig_{\psi''DD}\psi''^\mu
D_i^\dag\stackrel{\leftrightarrow}{\partial_\mu}
D_i \ , \\
\mathcal{L}_{\psi DD^*}&=&g_{\psi
DD^*}\epsilon_{\mu\nu\alpha\beta}\partial^\mu\psi_n^\nu\{D_i^{*\beta\dag}\stackrel{\leftrightarrow}{\partial^\alpha}D_i-
D_i^\dag\stackrel{\leftrightarrow}{\partial^\alpha}D_i^{*\beta}\} \ ,\\
\mathcal{L}_{D^*D\pi}&=&-ig_{D^*DP}\left(D^i\partial^\mu
P_{ij}D_\mu^{*j\dag}-D_\mu^{*i}\partial^\mu P_{ij}D^{j\dag}\right) \
,
\end{eqnarray}
where the coupling $g_{\psi DD^*}\equiv M_\psi/(f_\psi\sqrt{M_{D^*}
M_D})\ \mathrm{GeV}^{-1}$. The same convention has also been adopted
in Ref.~\cite{Wang:2012mf}.

The decay amplitude via the $D\bar{D}(D^*)$ loop is
\begin{eqnarray}
\mathcal{M}_{fi}&=&\sum_{Polarization}\int\frac{\mathrm{d}^4p_5}{(2\pi)^4}(2g_{\psi''DD}p_3\cdot
\epsilon_1)(-2g_{\psi
DD^*}\epsilon_{\mu\nu\alpha\beta}q_2^\mu\epsilon_2^{*\nu}
p_5^\alpha\epsilon_5^\beta)(\frac{-g_{D^*DP}}{\sqrt{2}}p\cdot\epsilon_5^*)\nonumber\\
&&\times\frac{i}{p_3^2-M_D^2}\frac{i}{p_4^2-M_D^2}\frac{i}{p_5^2-M_{D^*}^2}
F(p_i^2) \ .
\end{eqnarray}
At each vertex of the loop diagram, the off-shell effect or the
finite size effect should be taken into account by introducing the
form factor $F(p_i^2)$, which can be regarded as the extended
version of the local couplings in the original effective Lagrangian.
The form factor is also necessary for cutting off the ultraviolet
divergence in the loop integrals.

Applying the Cutkosky rule, the discontinuity of the decay amplitude
is
\begin{eqnarray}
 \emph{Disc.}\mathcal{M}_{fi}&=&-i(-2\pi
i)^2\int\frac{\mathrm{d}^4p_5}{(2\pi)^4}(2g_{\psi''DD})(p_3\cdot\epsilon_1)\epsilon_{q_2\epsilon_2^*p_5\mu}p_\nu\left(-g^{\mu\nu}+\frac{p_5^\mu
p_5^\nu}{M_{D^*}^2}\right)\nonumber\\
&&\times\delta(p_3^2-M_D^2)\times\delta(p_4^2-M_D^2)\times\frac{F_{\psi
DD^*}(p_5^2)F_{D^*D\pi^0}(p_5^2)}{p_5^2-M_{D^*}^2}
\end{eqnarray}
with
$\epsilon_{q_2\epsilon_2^*p_5\mu}\equiv\epsilon_{\alpha\beta\gamma\mu}q_2^\alpha\epsilon_2^{*\beta}p_5^\gamma$
for a short notation. Then, there are two form factors depending
only on the virtuality of $D^*$ left in our calculation. In
practice, the product of these form factors can be parameterized as
the product of the local couplings and an empirical form factor:
\begin{eqnarray}
F_{\psi_nDD^*}(p_5^2)F_{D^*D\pi^0}(p_5^2)\equiv
-2g_{\psi_nDD^*}\frac{-g_{D^*DP}}{\sqrt{2}}F(p_5^2) \ ,
\end{eqnarray}
where a dipole form factor is adopted,
\begin{eqnarray}\label{Eq:Dipole}
F(p_5^2)=\left(\frac{\Lambda^2-M_{D^*}^2} {\Lambda^2-p_5^2}\right)^2,
\end{eqnarray}
with $\Lambda=M_{D^*}+\alpha\Lambda_{QCD}$ and $\Lambda_{QCD}=0.22 \
\mathrm{GeV}$.

Further deduction gives the discontinuity of the decay amplitude
\begin{eqnarray}
\emph{Disc.}\mathcal{M}_{fi}&=&2ig_{\psi''DD}\int\frac{\mathrm{d}^3\vec{p}_3}{(2\pi)^32E_{p_3}2E_{p_4}}\frac{F_{\psi DD^*}(p_5^2)F_{D^*D\pi^0}(p_5^2)}{p_5^2-M_{D^*}^2}\nonumber\\
&&\times2\pi\delta(p_3^0+p_4^0-q_1^0)\epsilon_{q_2\epsilon_2^*p_3
p}p_3\cdot\epsilon_1|_{constraints} \ ,
\end{eqnarray}
where $E_{p_i}=\sqrt{|\vec{p}_i|^2+M_i^2}$, and $constraints\equiv
(p_3^0>0,p_4^0>0,p_3^2=M_D^2,p_4^2=M_D^2,p_5=q_2-p_3,\vec{p}_4=\vec{p}_5+\vec{p})$.

Extracting the antisymmetric Lorentz structure
$\epsilon_{q_2\epsilon_2^*\mu p}\epsilon_{1\nu}$, we get the tensor
amplitude
\begin{eqnarray}
\emph{Disc.}\mathcal{M}_{fi}^{\mu\nu}&=&2ig_{\psi''DD}\int\frac{\mathrm{d}^3\vec{p}_3}{(2\pi)^32E_{p_3}2E_{p_4}}\frac{F_{\psi DD^*}(p_5^2)F_{D^*D\pi^0}(p_5^2)}{p_5^2-M_{D^*}^2}\nonumber\\
&&\times2\pi\delta(p_3^0+p_4^0-q_1^0)p_3^\mu p_3^\nu|_{constraints}
\end{eqnarray}
From Lorentz invariance, this tensor structure can be
decomposed into terms built out of the external momenta and metric
tensor:
\begin{eqnarray}
\emph{Disc.}\mathcal{M}_{fi}^{\mu\nu}&=&g^{\mu\nu}\mathcal{M}_A+p^\mu
p^\nu\mathcal{M}_B+q_2^\mu q_2^\nu \mathcal{M}_C+(p^\mu
q_2^\nu+p^\nu q_2^\mu)\mathcal{M}_D,
\end{eqnarray}
where only $\mathcal{M}_A$ will contribute to the final result when
we contract the tensor amplitude with the extracted antisymmetric
Lorentz structure.

Contracting the tensor amplitude with the metric tensor, we obtain
\begin{eqnarray}
g_{\mu\nu}\emph{Disc.}\mathcal{M}_{fi}^{\mu\nu}&=&2ig_{\psi''DD}\int\frac{\mathrm{d}^3\vec{p}_3}{(2\pi)^32E_{p_3}2E_{p_4}}\frac{F_{\psi DD^*}(p_5^2)F_{D^*D\pi^0}(p_5^2)}{p_5^2-M_{D^*}^2}\nonumber\\
&&\times2\pi\delta(p_3^0+p_4^0-q_1^0)p_3^2|_{constraints}\nonumber\\
&=&2ig_{\psi''DD}\int\frac{|\vec{p}_3|^2\mathrm{d}|\vec{p}_3|\sin\theta\mathrm{d}\theta\mathrm{d}\varphi}{(2\pi)^32E_{p_3}2E_{p_4}}\frac{F_{\psi DD^*}(p_5^2)F_{D^*D\pi^0}(p_5^2)}{p_5^2-M_{D^*}^2}\nonumber\\
&&\times2\pi\delta(p_3^0+p_4^0-q_1^0)p_3^2|_{constraints}\nonumber\\
\end{eqnarray}
In the rest frame of $\psi(3770)$ and setting the direction of
$\vec{q}_2$ as the $z$-axis, the dynamic variables can be expressed
as
\begin{eqnarray}
h&\equiv&
|\vec{q}_2|=\sqrt{\left(\frac{M_\psi^2+M_{\psi''}^2-M_\pi^2}{2M_{\psi''}}\right)^2-M_\psi^2}\nonumber\\
\frac{r}{2}&\equiv&|\vec{p}_3|=\frac{\sqrt{M_{\psi''}^2-4M_D^2}}{2}\nonumber\\
v&\equiv&q_2^0=\frac{M_{\psi''}^2+M_\psi^2-M_\pi^2}{2M_{\psi''}}\nonumber\\
w&\equiv&p^0=\frac{M_{\psi''}^2-M_\psi^2+M_\pi^2}{2M_{\psi''}}\nonumber\\
q_1^0&=&v+w=M_{\psi''}\nonumber\\
p_3^0&=&\frac{v+w}{2}
\end{eqnarray}
Then, the Lorentz invariant amplitude is
\begin{eqnarray}
g_{\mu\nu}\emph{Disc.}\mathcal{M}_{fi}^{\mu\nu}&=&ig_{\psi''DD}\times\frac{r}{8\pi(v+w)}\times\int_{-1}^1\mathrm{d}x\,\frac{M_D^2
F_{\psi
DD^*}(x)F_{D^*D\pi^0}(x)}{G(x)}\nonumber\\
\end{eqnarray}
where the propagator of $D^*$ is $G(x)=M_\psi^2+M_D^2-2[v(v+w)/2-h r
x/2]-M_{D^*}^2$.

Similarly, we can get other three Lorentz invariant amplitudes:
\begin{eqnarray}
p_\mu
p_\nu\emph{Disc.}\mathcal{M}_{fi}^{\mu\nu}&=&ig_{\psi''DD}\times\frac{r}{8\pi(v+w)}\nonumber\\
&&\times\int_{-1}^1\mathrm{d}x\,\frac{[w(v+w)/2+h r x/2]^2 F_{\psi
DD^*}(x)F_{D^*D\pi^0}(x)}{G(x)}\nonumber\\
\end{eqnarray}

\begin{eqnarray}
q_{2\mu}
q_{2\nu}\emph{Disc.}\mathcal{M}_{fi}^{\mu\nu}&=&ig_{\psi''DD}\times\frac{r}{8\pi(v+w)}\nonumber\\
&&\times\int_{-1}^1\mathrm{d}x\,\frac{[v(v+w)/2-h r x/2]^2 F_{\psi
DD^*}(x)F_{D^*D\pi^0}(x)}{G(x)}\nonumber\\
\end{eqnarray}

\begin{eqnarray}
p_\mu
q_{2\nu}\emph{Disc.}\mathcal{M}_{fi}^{\mu\nu}&=&ig_{\psi''DD}\times\frac{r}{8\pi(v+w)}\nonumber\\
&&\times\int_{-1}^1\mathrm{d}x\,\frac{[v(v+w)/2-h r x/2][w(v+w)/2+h
r x/2] F_{\psi
DD^*}(x)F_{D^*D\pi^0}(x)}{G(x)}\nonumber\\
\end{eqnarray}

Solving these four equations simultaneously, we can get the
complicated expression of the invariant amplitude $\mathcal{M}_A$
and the absorptive part of the decay amplitude
$\emph{Disc.}\mathcal{M}_{fi}$. The charge conjugate contribution
gives the same result.

Because of the mass of $\psi(3770)$ being above the charmed meson
pair, the coupling constant $g_{\psi''DD}$ and the isospin
difference may be difficult to get from theory because of the
rescattering mechanism. So we will extract this coupling directly
from the experimental data:
\begin{eqnarray}
\Gamma_{\psi''\rightarrow
D\bar{D}}&=&\frac{4g_{\psi''DD}^2|\vec{p}_3|}{8\pi
M_{\psi''}^2}\times\frac{1}{3}\sum_{\epsilon_1}
(p_3\cdot\epsilon_1)(p_3\cdot\epsilon_1^*)\nonumber\\
&=&\frac{g_{\psi''DD}^2|\vec{p}_3|}{6\pi
M_{\psi''}^2}\left(-M_D^2+\frac{M_{\psi''}^2}{4}\right) \ .
\end{eqnarray}

\subsection{Dispersive part}\label{sec:1.2}

In principle, all the meson loops of which the thresholds are above
the $\psi(3770)$ mass would contribute to the dispersive part (i.e.
the real part) of the transition amplitude. Because of the
introduction of form factors in the loop integrals, some model
dependence seems inevitable in the evaluation of the real part.
Given that the imaginary part of the amplitude can be reliably
determined as in the previous subsection, we shall apply the
dispersion relation to obtain the real part of the decay amplitude.
Taking the assumption that the spectral density can be approximated
by the extrapolation $\mathcal{M}_A(M_{\psi''}^2)\rightarrow
\mathcal{M}_A(s_1)$, we have the unsubtracted dispersion relation:
\begin{eqnarray}
Re[\mathcal{M}_{fi}^{Tot}]=\frac{1}{2\pi
i}\left(\mathcal{P}\int_{(2M_{D^+})^2}^{{th_C}}\frac{2\mathcal{M}_A^{C}(s_1)}{s_1-M_{\psi''}^2}\mathrm{d}s_1
+\mathcal{P}\int_{(2M_{D^0})^2}^{{th_N}}\frac{2\mathcal{M}_A^{N}(s_1)}{s_1-M_{\psi''}^2}\mathrm{d}s_1\right)
\ , \label{eq:1.2.1}
\end{eqnarray}
where $\mathcal{M}_A^{C/N}$ corresponds to the charged or neutral
$D$ meson loop's contribution, and the factor $2$ in front of
$\mathcal{M}_A(s_1)$ refers to the charge conjugate contribution.
Then the total decay width is
\begin{eqnarray}
\Gamma&=&\frac{h}{8\pi
M_{\psi''}^2}\int\frac{\mathrm{d}\Omega_{cm}}{4\pi}
\left[Re[\mathcal{M}_{fi}^{Tot}]^2+\left(\frac{2\mathcal{M}_A^{Tot}}{2i}\right)^2\right]
\times\frac{1}{3}\sum_{\epsilon_1,\epsilon_2}\epsilon_{q_2\epsilon_2^*\epsilon_1p}
\epsilon_{q_2\epsilon_2\epsilon_1^*p}\nonumber\\
&=&-\frac{h}{12\pi
M_{\psi''}^2}\left[Re[\mathcal{M}_{fi}^{Tot}]^2+\left(\frac{2\mathcal{M}_A^{Tot}}{2i}\right)^2\right]\nonumber\\
&&\times \left[M_\pi^2\left(v-w+\frac{M_\pi^2}{v+w}\right)(v+w)-(v
w+h^2)^2\right] \ ,
\end{eqnarray}
where $\mathcal{M}_A^{Tot}=\mathcal{M}_A^{C}+\mathcal{M}_A^{N}$.

Two points should be stressed: one is the upper limit of the
dispersive integral, and the other is the virtuality dependence of
the coupling $g_{\psi''DD}$. Generally speaking, the upper limit of
the dispersive integral should be infinity from the mathematical
viewpoint. But in practice, we only take a finite effective
threshold ${th}$ because the spectral density is only an
approximation.  It is presented that the form factor of
$D+J/\psi(virtual)\rightarrow D$ is harder than that of
$D+\rho(virtual)\rightarrow D$ because $D$ can ``see'' the size of
smaller $J/\psi$~\cite{Matheus:2002nq}. We expect that the heavier
$\psi(3770)$ also gives a harder $g_{\psi''DD}(s_1)$ form factor at
large $s_1$ so that in a limited $s_1$ region the $s_1$-dependence
can be neglected. There is also literature \cite{Liu:2009dr} to take
this $s_1$-dependence into account by adding a suppression factor
$\exp(-I|\vec{p}_3|^2)$ into the integrand of Eq.~\eqref{eq:1.2.1},
where $I$ is the square of the interaction
length~\cite{Pennington:2007xr}. We will discuss both points in
detail in the following numerical analysis.
\section{Form factors for the off-shell vertex couplings}\label{sec:2}
\subsection{\boldmath QSSR  reanalysis of the form factor $F_{\psi DD^*}(p_5^2)$\unboldmath}\label{sec:2.1}
Since the mass of $J/\psi$ is below the lowest threshold of open
charm $D\bar{D}$, it is not possible to measure the form factor
$F_{\psi DD^*}(p_5^2)$ in experiment directly. There has been a
systematic investigation of the charmonium to open charmed meson
form factors in the framework of
QSSR~\cite{RodriguesdaSilva:2003hh,Matheus:2003pk}. As a crucial
criterion of QSSR, the pole contribution should take a dominant part
in the dispersion integral. To our surprise, it seems not possible
to satisfy this condition with the parameters given in the
literature. This stimulate us to reinvestigate the $F_{\psi
DD^*}(p_5^2)$ with the improved QSSR approach, and crosscheck the
result with finite energy sum rules (FESR).

We shall be concerned with the three-point correlation function:
\begin{eqnarray}
\Gamma_{\mu\nu}(q_2,p_3)=\int\mathrm{d}^4x\,\mathrm{d}^4y
e^{ip_3\cdot x}e^{-i(p_3-q_2)\cdot y}\langle 0|T\left\{J^3(x)
J_\mu^{2\dag}(y)J_\nu^{1\dag}(0)\right\}|0\rangle \ ,
\end{eqnarray}
where $J_\nu^1=\bar{c}\gamma_\nu c$, $J_\mu^2=\bar{q}\gamma_\mu c$
and $J^3=i\bar{q}\gamma_5 c$ denote the interpolating currents for
the incoming $J/\psi(q_2,\epsilon_2)$, incoming
$D^*(p_5,\epsilon_5)$ and outgoing $D$, respectively. Taking the
advantage of the unique Lorentz structure for the $VVP$ coupling, we
can decompose $\Gamma_{\mu\nu}$ simply as:
\begin{eqnarray}
\Gamma_{\mu\nu}(q_2,p_3)\equiv
\Lambda(q_2^2,p_3^2,p_5^2)\epsilon_{\mu\nu\alpha\beta}q_2^\alpha
p_5^\beta \ ,
\end{eqnarray}
where $p_3=p_5+q_2$. The above expression has an arbitrary sign
compared with that the preceding section. Using a double dispersion
relation, one can express the invariant amplitude as:
\begin{eqnarray}
\Lambda(q_2^2,p_3^2)=\frac{-1}{4\pi^2}\int\mathrm{d}s\,\mathrm{d}u\frac{\rho(s,u,p_5^2)}{(s-q_2^2)(u-p_3^2)}
\ .
\end{eqnarray}
For the $D^*$-meson off-shell, the spectral density can be obtained
from the Cutkosky rule presented in the previous section.

On the phenomenological side, the three-point correlation function
can be approximated by the lowest resonance plus the ``QCD
continuum'' contributions, where the latter come from the
discontinuity of the QCD diagrams from a threshold:
\begin{eqnarray}
\sqrt{u_0}(\sqrt{s_0})\equiv M_D (M_\psi) +\Delta \ ,
\end{eqnarray}
and smears the contributions of all higher resonance contributions.
In this way, the phenomenological part of the three-point function
reads:
\begin{eqnarray}
\Lambda^{phen}=\frac{\delta_c F_{\psi DD^*}(p_5^2)\epsilon_{\mu\nu
q_2p_5}}{(p_5^2-M_{D^*}^2)(q_2^2-M_\psi^2)(p_3^2-M_D^2)}+\textrm{``QCD\,\,
continuum''}
\end{eqnarray}
where $\delta_c\equiv {M_D^2M_{D^*}M_\psi f_D f_{D^*}
f_\psi}/{m_c}$, the form factor with virtual $D^*$ is defined as
\begin{eqnarray}
\langle D(p_3)|J_\mu^{2\dag}|J/\psi(q_2)\rangle&=& \frac{ \langle
D^*(p_5)|J_\mu^{2\dag}|0\rangle \langle
D(p_3)|D^*(p_5)J/\psi(q_2)\rangle}{p_5^2-M_{D^*}^2}\ ,
\end{eqnarray}
with
\begin{eqnarray}
\langle D(p_3)|D^*(p_5)J/\psi(q_2)\rangle&\equiv& F_{\psi
DD^*}(p_5^2)\epsilon_{\epsilon_2\epsilon_5p_3p_5} \ ,
\end{eqnarray}
and the decay constants are normalized as:
\begin{eqnarray}
\langle D^*(p_5)|J_\mu^{2\dag}|0\rangle&=&M_{D^*}f_{D^*}\epsilon_\mu^* \ ,\nonumber\\
\langle 0|J^3|D(p_3)\rangle&=&\frac{M_D^2 f_D}{m_c} \ ,\nonumber\\
\langle J/\psi(q_2)|J_\nu^1|0\rangle&=&M_\psi f_\psi \epsilon_\nu^*
\ .
\end{eqnarray}

Matching the two sides of correlation function, and performing the
Borel transformation (Laplace SR), the lowest perturbative diagram
gives
\begin{eqnarray}
F_{\psi
DD^*}(p_5^2)&=&-\frac{1}{4\pi^2}\frac{p_5^2-M_{D^*}^2}{\delta_c}\int_{4m_c^2}^{s_0}\int_{u_{min}}^{u_0}\mathrm{d}s\,\mathrm{d}u \,\,\rho(u,s,t) \nonumber\\
&&\times e^{-(s-M_\psi^2)\tau_1}
e^{-(u-M_D^2)\tau_2}\theta(u_{max}-u) \ ,
\end{eqnarray}
with
\begin{eqnarray}
  \rho(s,t,u)&=&\frac{3m_c}{\sqrt{\lambda}}\left(1+\frac{s\lambda_2}{\lambda}\right) \ ,\nonumber\\
  u_{min}^{max}&=&\frac{1}{2m_c}\left[-s t+m_c^2(s+2t)\pm
  \sqrt{s(s-4m_c^2)(t-m_c^2)^2}\right] \ ,
\end{eqnarray}
where $t=p_5^2$, $\lambda\equiv(u+s-t)^2-4u s$, $\lambda_2\equiv
u+t-s+2m_c^2$, and $\tau_{1,2}$ are the inverse squares of the
corresponding Borel masses. Taking the limits $\tau_1\rightarrow 0$
and $\tau_2\rightarrow 0$, we obtain the FESR. Here, we neglect the
numerically small gluon condensate
contribution~\cite{Matheus:2003pk}.

\begin{figure}[!htp]
\centering
\subfigure[]{\label{fig:mini:FD:a}
\includegraphics[width=.49\textwidth]{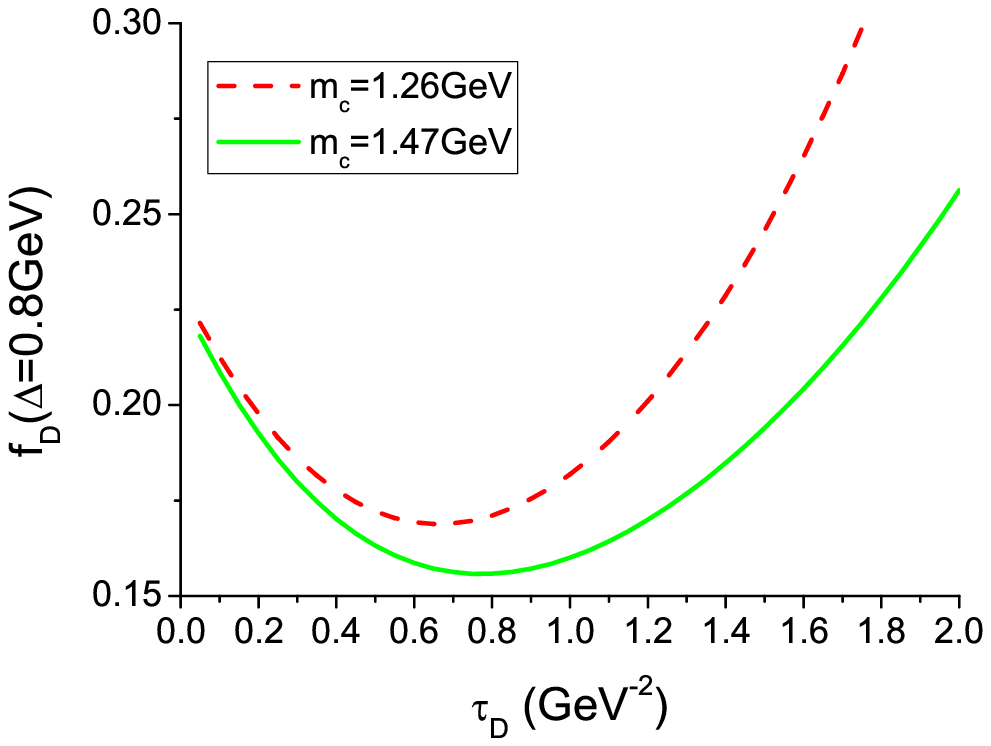}}
\subfigure[]{\label{fig:mini:FD:b}
\includegraphics[width=.49\textwidth]{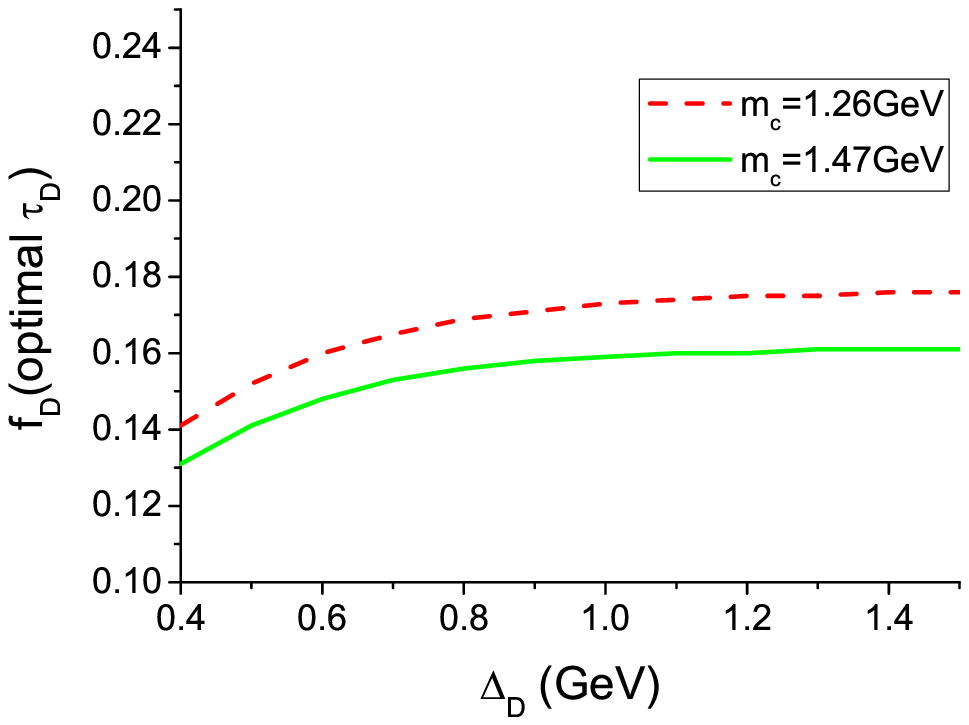}}
\caption{(color online). (a) The two-point SR of $f_D$ versus
$\tau_D$ with $\Delta_D=0.8 \ \mathrm{GeV}$. The red dashed line is
for $m_c=1.26 \ \mathrm{GeV}$ and the green solid line for $m_c=1.47
\ \mathrm{GeV}$. (b) $f_D$ versus $\Delta_D$ with the minimum of
$\tau_D$ adopted. } \label{fig:FD}
\end{figure}

\begin{figure}[!htp]
\centering
\subfigure[]{\label{fig:mini:FJ:a}
\includegraphics[width=.49\textwidth]{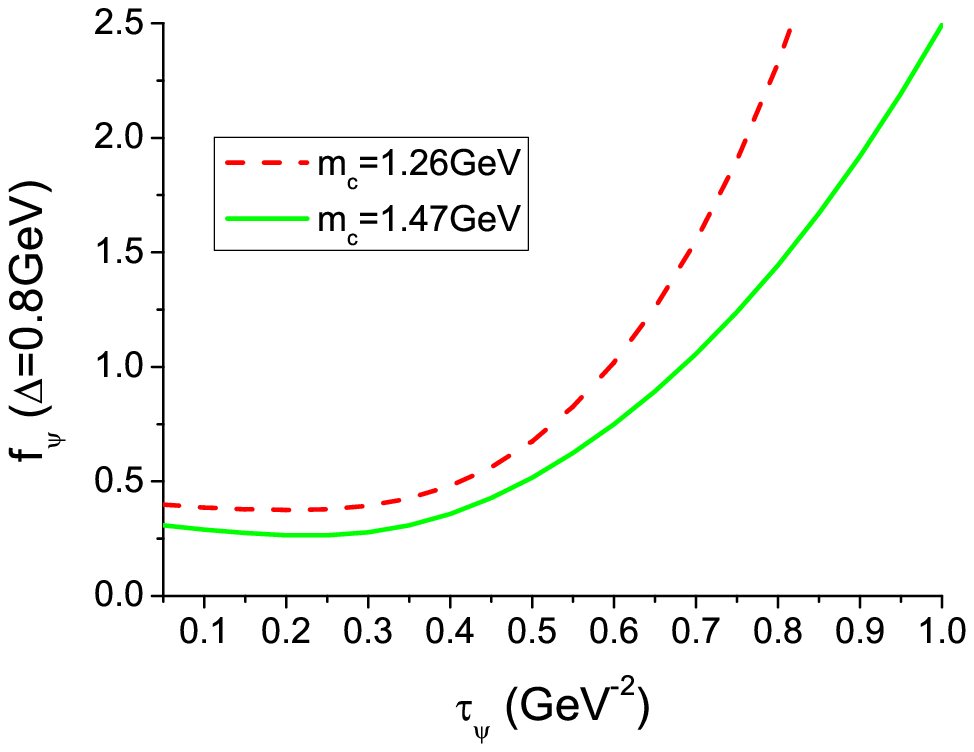}}
\subfigure[]{\label{fig:mini:FJ:b}
\includegraphics[width=.49\textwidth]{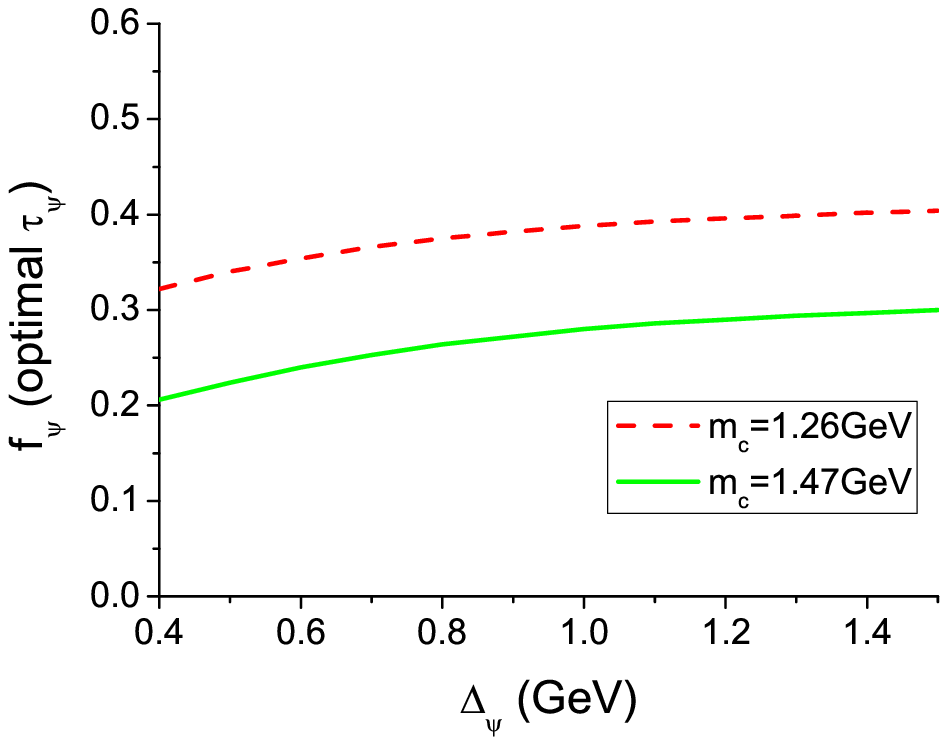}}
\caption{(color online). (a) The two-point SR of $f_\psi$ versus
$\tau_\psi$ with $\Delta_\psi=0.8 \ \mathrm{GeV}$. The red dashed
line is for $m_c=1.26 \ \mathrm{GeV}$ and the green solid line for
$m_c=1.47 \ \mathrm{GeV}$. (b) $f_\psi$ versus $\Delta_\psi$ with
the minimum of $\tau_\psi$ adopted.} \label{fig:FJ}
\end{figure}

\begin{figure}[!htp]
\centering
\includegraphics[width=0.65\textwidth]{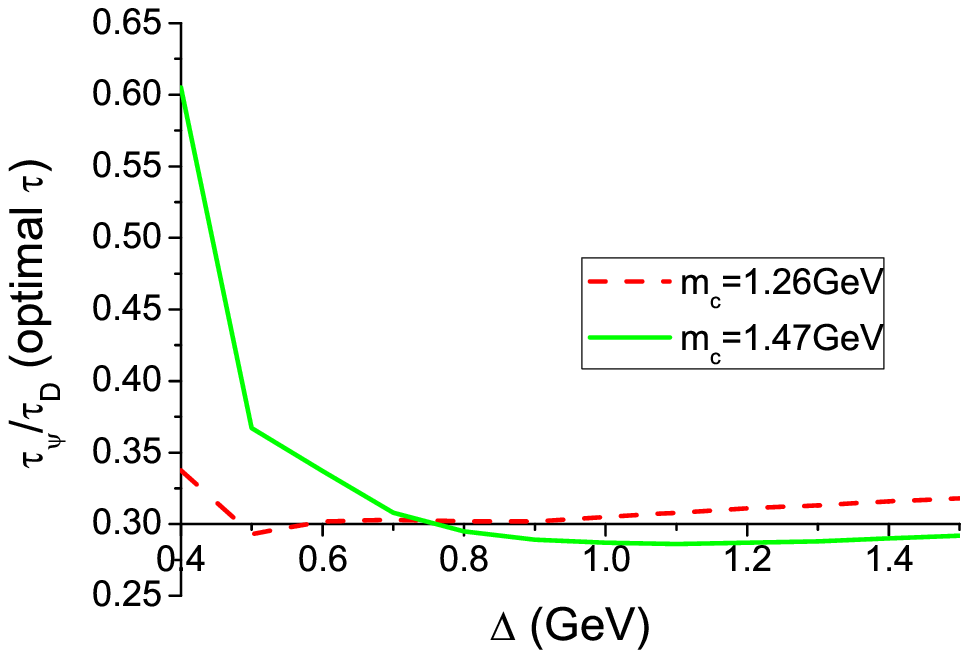}
\caption{(color online). Behavior of the ratio of SR variables
$\tau_\psi/\tau_D$ versus $\Delta$ for two values of $m_c=1.26 \
\mathrm{GeV}$ (red dashed line) and $1.47 \ \mathrm{GeV}$ (solid
green line).} \label{fig:Fraction}
\end{figure}

To the leading order approximation where the three-point correlation
function is evaluated, it is consistent to extract the decay
constants $f_D$ and $f_\psi$ from the corresponding two-point
functions at the lowest order other than the value extracted  from
the experiment directly, e.g. $f_\psi=0.405\pm 0.015 \
\mathrm{GeV}$. The QCD expressions of the pseudoscalar and vector
two-point functions are well known~\cite{TwoPointSR}. We show our
analysis for $f_D$ and $f_\psi$ in Fig.~\ref{fig:FD} and
Fig.~\ref{fig:FJ}, respectively. Stabilities in both the two-point
sum rule variables $\tau_{\psi,D}$ and variation of the continuum
threshold $\Delta$ are observable. We show in
Fig.~\ref{fig:Fraction} the $\Delta$ behavior of the ratio
$\tau_\psi/\tau_D$ which is rather stable, especially for $m_c=1.26
\ \mathrm{GeV}$. The obtained optimal ratios (stable with $\Delta$)
are:
\begin{equation}
  \frac{\tau_\psi}{\tau_D} \simeq
\begin{cases}
  0.30&  \mathrm{for}\ m_c=1.26 \ \mathrm{GeV},\\
  0.28&  \mathrm{for}\ m_c=1.47 \ \mathrm{GeV},
\end{cases}
\end{equation}
while the ad-hoc phenomenological choice used in the literature is:
\begin{eqnarray}
\frac{\tau_\psi}{\tau_D}=\frac{M_D^2}{M_\psi^2}=0.364 \ .
\end{eqnarray}
The relations between the two-point parameters $\tau_{\psi,D}$ and
the corresponding three-point parameters $\tau_{1,2}$
are~\cite{Dosch:1997zx}:
\begin{eqnarray}
  \tau_1\simeq\frac{\tau_\psi}{2}, \,\,
 \tau_2\simeq\frac{\tau_D}{2} \ .
\end{eqnarray}

\begin{figure}[!htp]
\centering \subfigure[]{\label{fig:mini:Ex:F3}
\includegraphics[width=.49\textwidth]{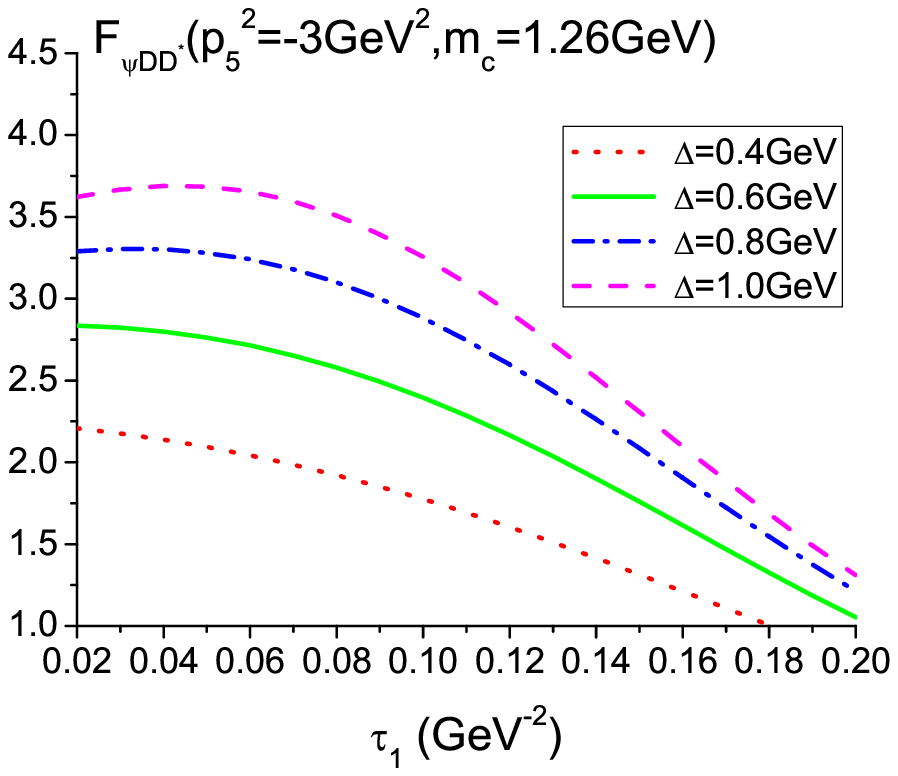}}
\subfigure[]{\label{fig:mini:Ex:R3}
\includegraphics[width=.49\textwidth]{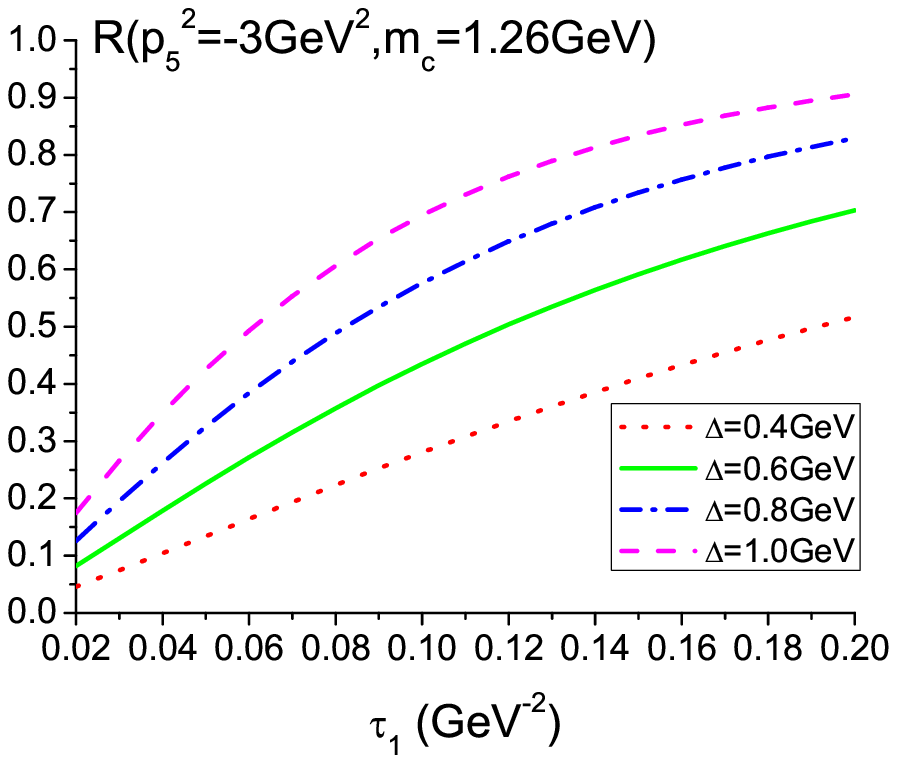}}
\caption{(color online). (a) The $\tau_1$ windows of the form factor
and (b) the contribution from the corresponding pole at $p_5^2=-3 \
\mathrm{GeV^2}$, where $m_c=1.26 \ \mathrm{GeV}$,
$\tau_1/\tau_2=0.3$, and $\Delta$ as a variable.} \label{fig:Ex}
\end{figure}

As follows, we adopt $m_c=1.26 \ \mathrm{GeV}$ and
$\Delta_D=\Delta_\psi\equiv\Delta$ as inputs since it is difficult
to find a global maximum with $m_c=1.47 \ \mathrm{GeV}$ in our
three-point SR and the variation of $\tau$ ratio is more stable with
$m_c=1.26 \ \mathrm{GeV}$\,~\cite{ft1}. To obtain more concrete
information about the form factor, we consider a large virtuality
interval $0\leq -p_5^2\leq 5 \ \mathrm{GeV^2}$, which is the same as
in Ref.~\cite{RodriguesdaSilva:2003hh}. As an illustration, we show
in Fig.~\ref{fig:Ex} the form factor $F_{\psi DD^*}(p_5^2)$ at
$p_5^2=-3 \ \mathrm{GeV^2}$ with both $\Delta\geq 0.4 \
\mathrm{GeV}$ and ${\tau_1}/{\tau_2}=0.3$. For simplicity,
$f_{D^*}=0.24 \ \mathrm{GeV}$ is the same as in
Ref.~\cite{RodriguesdaSilva:2003hh}. The ratios of the pole
contribution versus the whole dispersion integral are also depicted
in Fig.~\ref{fig:Ex} and parameterized as
\begin{eqnarray}
\mathrm{R}&\equiv &\mathrm{\frac{PI}{WI}},\\
\mathrm{PI}&=&\int_{4m_c^2}^{s_0} \mathrm{d} s\int_{m_c^2}^{u_0}
\mathrm{d} u\,
\rho(s,u,Q^2)\nonumber\\
&&\times\theta(u_\mathrm{max}-u)\theta(u-u_\mathrm{min})e^{-s \tau_1-u \tau_2},\\
\mathrm{WI}&=&\int_{4m_c^2}^{\infty} \mathrm{d}
s\int_{u_\mathrm{min}}^{u_\mathrm{max}} \mathrm{d} u\,
\rho(s,u,Q^2)\nonumber\\
&&\times\theta(u_\mathrm{max}-u)\theta(u-u_\mathrm{min})e^{-s
\tau_1-u \tau_2}. \label{eq:1.9}
\end{eqnarray}

\begin{figure}[!htp]
\centering \subfigure[]{\label{fig:mini:FF:a}
\includegraphics[width=.49\textwidth]{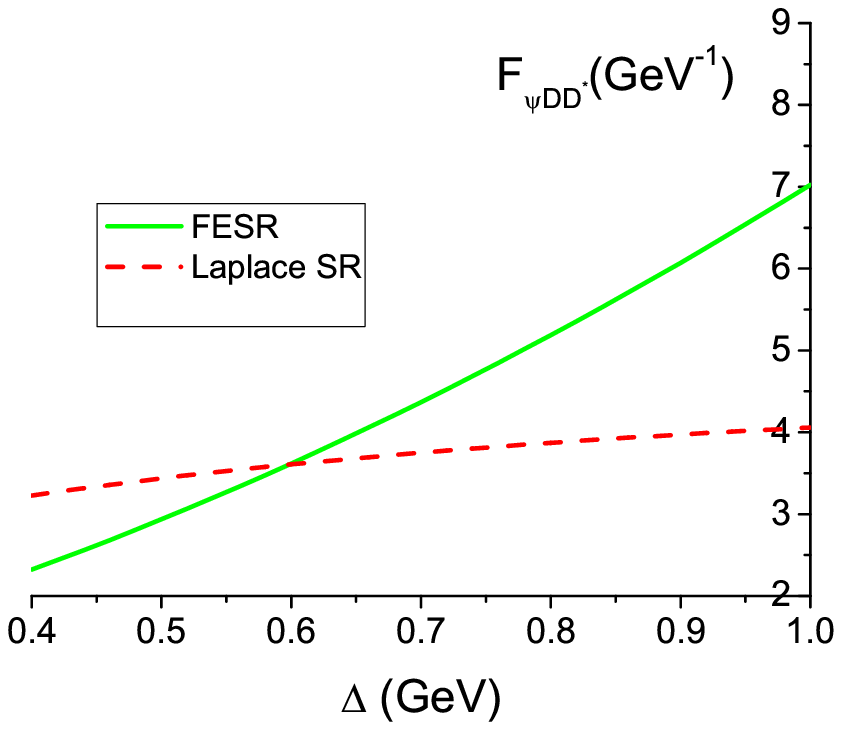}}
\subfigure[]{\label{fig:mini:FF:b}
\includegraphics[width=.49\textwidth]{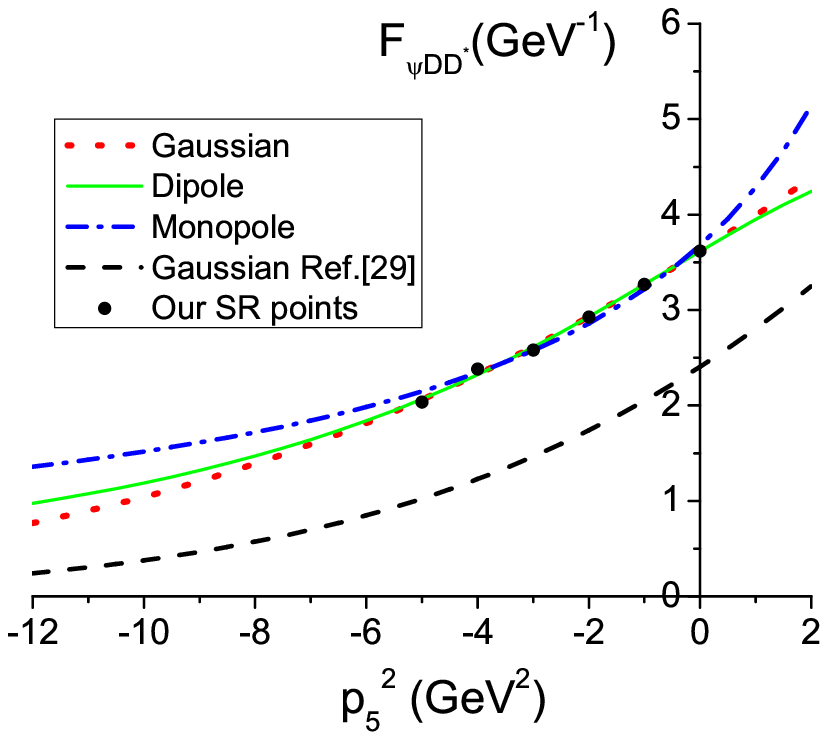}}
\caption{(color online). (a) The intersection point of Laplace SR
and FESR at $p_5^2=0\mathrm{GeV}^2$. (b) The $J/\psi DD^*$ form
factor derived from our method compared with the Gaussian fit
~\cite{RodriguesdaSilva:2003hh}. The blue dash-dotted line is our
monopole fit, the green solid line is our dipole fit, the red dotted
one is our Gaussian fit, and the black dashed one is the Gaussian
fit ~\cite{RodriguesdaSilva:2003hh}. The dots are our SR result.}
\label{fig:FF}
\end{figure}

With the increase of $\Delta$, the pole contributions will become
larger. It is obvious to see that the pole contributions are less
than one half at the maximum $\tau_1=0.05 \ \mathrm{GeV}$ even with
$\Delta$ as large as $1 \ \mathrm{GeV}$. The situation will be worse
with larger $D^*$ virtuality. This phenomenon seems to be a common
problem for the form factors of charmonium to open charmed mesons.
To avoid this difficulty of the SR criterion, we deduce the form
factors from Laplace SR varying with different $\Delta$, and also
show the predictions from FESR in terms of $\Delta$. In principle,
these two SRs should give the same solution, which means that the
result at the intersection point is the reliable one from QSSR, see
Fig.~\ref{fig:mini:FF:a}. The form factor from the above method is
shown in Fig.~\ref{fig:mini:FF:b}, and we use three different
parameterizations to extend the form factor to broader regions of
the $D^*$ virtuality:

\begin{equation}
F_{\psi DD^*}(p_5^2)=\begin{cases}
10.58 \exp\left[-\genfrac{}{}{}{0}{(-p_5^2+21.30)^2}{422.62}\right]&\text{Gaussian}\\[10pt]
\genfrac{}{}{}{0}{332.61}{p_5^4-8.79 p_5^2+91.98} & \text{Dipole}\\[10pt]
\genfrac{}{}{}{0}{-25.83}{p_5^2-7.03} &  \textrm{Monopole}
                     \end{cases}
\end{equation}

The form factor obtained in Ref.~\cite{RodriguesdaSilva:2003hh} with
fixed $f_\psi =0.405 \ \mathrm{GeV}$, $f_D=0.17 \ \mathrm{GeV}$,
$m_c=1.3 \ \mathrm{GeV}$, $\Delta_\psi=\Delta_D=0.5 \ \mathrm{GeV}$
and $0.09<\tau_1<0.14 \ \mathrm{GeV^{-2}}$ are parameterized by the
Gaussian formula~\cite{RodriguesdaSilva:2003hh}
\begin{eqnarray}
F_{\psi DD^*}(p_5^2)=19.9 \exp
\left[\frac{-(-p_5^2+27)^2}{345}\right]\,\, .
\end{eqnarray}
Below the $D^*$ threshold, our improved form factors are slightly
larger and decline slower than the one in
Ref.~\cite{RodriguesdaSilva:2003hh}. In fact, the form factors used
in our following calculation are usually restricted to a small
region $-5<p_5^2<2 \ \mathrm{GeV^2}$ with on-shell $D$ mesons.
Therefore, those different parameterizations would not bring
noticeable differences to the calculation results, although in a
broader momentum region they turn out to be different from each
other especially in the timelike region. Usually, pQCD predicts the
power falloff of the form factors, we will use the dipole fit in the
following calculations. Notice that there is no real roots, i.e. the
unphysical state, in the denominator of our dipole fit (we label it
as power fit hereafter to distinguish it from the empirical dipole
form factor).

\boldmath\subsection{The Form factor of $F_{D^*D\pi^0}(p_5^2)$}\unboldmath\label{sec:2.2}
As mentioned earlier, the form factor $F_{D^*D\pi^0}(p_5^2)$ is
determined by the $D$ meson semileptonic decays, i.e. $D\to \pi^0
l\nu$. In the momentum transfer region $p_5^2<2 \ \mathrm{GeV}^2$,
we expect that the $D^*$ pole has the dominant contribution. Thus,
the transition matrix element can be expressed as

\begin{eqnarray}
 &&\langle
\pi^0(p)|\bar{d}\gamma_\mu
  c|D^+(p_4)\rangle \nonumber\\
&\sim&\sum_{\epsilon_5}  \langle
\pi^0(p)D^{*+}(p_5)|D^+(p_4)\rangle\langle
  0|\bar{d}\gamma_\mu
  c|D^{*+}(p_5)\rangle\frac{1}{p_5^2-M_{D^{*+}}^2}\nonumber\\
  &=&\frac{F_{D^*D\pi^0}(p_5^2)M_{D^{*+}}f_{D^{*+}}}{p_5^2-M_{D^{*+}}^2}
  \left(-p_\mu+\frac{p_{5\mu} p\cdot p_5}{M_{D^{*+}}^2}\right) \ ,
\end{eqnarray}
where we have used the following definitions consistent with the
effective Lagrangian:
\begin{eqnarray}
 \langle
\pi^0(p)D^{*+}(p_5)|D^+(p_4)\rangle&=&F_{D^*D\pi^0}(p_5^2)(p\cdot\epsilon_5^*) \ ,\nonumber\\
\langle
  0|\bar{d}\gamma_\mu
  c|D^{*+}(p_5)\rangle&=&M_{D^{*+}}f_{D^{*+}}\epsilon_{5\mu} \ .
\end{eqnarray}

The transition matrix element of the weak decay can be defined
as~\cite{Khodjamirian:2009ys}
\begin{eqnarray}
\langle\pi^0(p)|\bar{d}\gamma_\mu c|D^+(p+p_5)\rangle&=&
\frac{1}{\sqrt{2}}[(2p+p_5)_\mu f_+(p_5^2)+p_{5\mu}f_-(p_5^2)] \ ,
\end{eqnarray}
where the form factor $f_+(p_5^2)$ has been measured with high
accuracy, and can be parameterized as a modified pole
formula~\cite{Besson:2009uv}
\begin{eqnarray}
f_+(p_5^2)=\frac{-f_+(0)M_{D^*}^2}{(p_5^2-M_{D^*}^2)\left(1-\alpha_0\frac{p_5^2}{M_{D^{*}}^2}\right)}
\ .
\end{eqnarray}

Compared with the $p^\mu$ part of the weak decay form factor
definition, we obtain the needed form factor:
\begin{eqnarray}
F_{D^*D\pi^0}(p_5^2)=\frac{\sqrt{2}f_+(0)M_{D^*}}{\left(1-\alpha_0
\frac{p_5^2}{M_{D^{*}}^2}\right)f_{D^{*}}} \ ,
\end{eqnarray}
where $\alpha_0=0.21$ for $D^0$, $\alpha_0=0.24$ for
$D^+$~\cite{Besson:2009uv}, and $f_+(0)=0.64$ from the lattice QCD
simulations~\cite{Aubin:2004ej,Bernard:2009ke} for our numerical
calculation. Note that $f_+(0)$ from QSSR~\cite{fplusSR} are
consistent with the lattice result very well. The local coupling
$g_{D^{*+}D^+\pi^0}$ can thus be extracted from the form factor at
$p_5^2=M_{D^*}^2$ with $f_{D^{*}}=0.24\mathrm{GeV}$, i.e.
\begin{eqnarray}
g_{D^*D\pi^0}(p_5^2)\equiv F_{D^*D\pi^0}(p_5^2=M_{D^*}^2)\simeq
9.97,
\end{eqnarray}
which is slightly different from the value extracted from the decay
of $D^*\rightarrow D+\pi$, i.e.
$g_{D^*D\pi^0}=g_{D^*DP}/\sqrt{2}=17.9/\sqrt{2}=12.7$~\cite{AA:2001}.
One can also extract this coupling from QSSR or QCD light-cone SR.
However, both SRs suffer from their inherent uncertainties and the
corresponding couplings from most SRs are nearly the same as what we
derived from the weak decay form factor. One can refer to
Ref.~\cite{Bracco:2011pg} for a review on this issue. Another reason
for the discrepancy of the coupling values is that in the momentum
transfer region $0<p_5^2<3 \ \mathrm{GeV}^2$ which corresponds to
the experimental kinematics, the form factor may vary drastically
near the pole position of $D^*$.

\begin{figure}[!htp]
\centering
\includegraphics[width=0.6\textwidth]{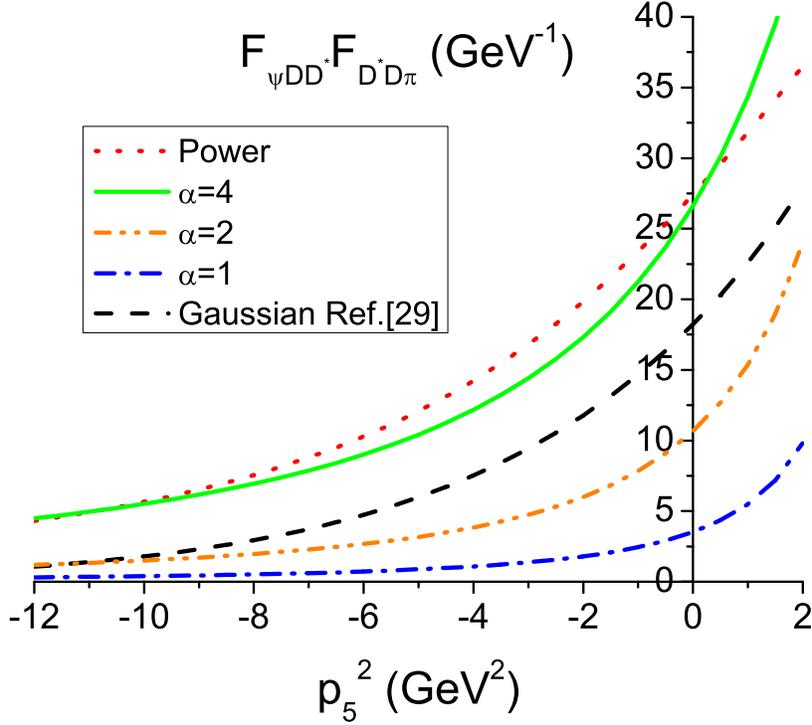}
\caption{(color online). The  QCD induced form factor $F_{\psi
DD^*}(p_5^2)F_{D^*D\pi^0}(p_5^2)$ (charged $D$) compared with the
empirical dipole form factors. The red dotted line is our power fit,
the black dashed line is the form factor obtained in
Ref.~\cite{RodriguesdaSilva:2003hh}, the green solid line, the
orange dash-dot-dotted line and the blue dash-dotted line correspond
to the empirical dipole form factor with $\alpha=4$, $\alpha=2$ and
$\alpha=1$, respectively.} \label{fig:Dipole}
\end{figure}

One could of course calculate the dispersive part with empirical
form factors, but we must emphasize that in some diagrams containing
more $D^*$ mesons, e.g. $D^*D(D^*)$ loop in the vector charmonium
decay to a $VP$ final state, the empirical dipole form factor used
by most of the references is not enough to suppress the ultraviolet
divergence in the loops so that we need other more complicated form
factors such as the Gaussian form factor. Thus, we leave the direct
calculation with empirical form factor aside. We present the
empirical dipole form factor (Eq.~(\ref{Eq:Dipole})) with different
cutoff in Fig.~\ref{fig:Dipole} and compare it  with our
QCD-motivated form factors. It is obvious that our form factors
favor $\alpha>2$, which is consistent with the value used in the
study of $X(3872)$ decays~\cite{Meng:2007cx}.

\section{Numerical results and discussion }\label{sec:3}
The determination of the effective threshold ${th}$ of the
dispersive part is not a trivial task. If we know the full
information of the spectral density, ${th}$ should be extended to be
infinity. Usually, ${th}\equiv (M_D+M_{D^*})^2$ is taken in the
literatures as a natural cutoff~\cite{Meng:2007cx} on the assumption
that the spectral density can be approximated by the extrapolation
of the imaginary part. As an improvement, it is assumed that the
corresponding effective threshold of $\psi(3770)\rightarrow
J/\psi+\eta$ containing  $u,d$ components should be the same as that
of $\psi(3770)\rightarrow J/\psi+\pi^0$. So, we can determine the
effective threshold from the decay width of $\psi(3770)\rightarrow
J/\psi+\eta$. The flavor mixing scheme is taken in our calculation,
and the mixing angle between
$|\bar{n}n\rangle\equiv(|\bar{u}u\rangle+|\bar{d}d\rangle)/\sqrt{2}$
and $|\bar{s}s\rangle$ is $\alpha_P\equiv
\theta_P+\arctan\sqrt{2}\simeq 38^\circ$, where $\theta_P$ is the
mixing angle between the flavor singlet and octet. Note that, as
mentioned in the Introduction that the IML contributions are
relatively enhanced by $1/v$ in comparison with the tree-level
contribution, it leads to the dominance of the IML contributions in
$\psi(3770)\to J/\psi+\eta$ which is consistent with the study of
Ref.~\cite{Zhang:2009kr}. It can be understood that the isospin
violation from the $\eta$-$\pi^0$ mixing is different from the IML
transitions. In the latter, there is no $\eta$ pole contributions to
the strong isospin violations.

In the following calculation of the dispersive part, we will take
the SU(3) flavor symmetry for the production of $s\bar{s}$ and
$q\bar{q}$ component within the light pseudoscalar mesons. It means
the dispersive part of $s\bar{s}$ is approximated by the average of
the dispersive integrals of $q\bar{q}$.

The decay constants of $g_{\psi''DD}$ in our numerical simulation
are listed in Table~\ref{tab:1}, where $g_{\psi''DD}^{low,cen,up}$
correspond to the lower bound, central value, and upper bound
allowed by the experimental data~\cite{Nakamura:2010zzi}.
%
From the experimental branching ratio $Br(\psi(3770)\rightarrow
J/\psi+\eta)=9\pm4\times 10^{-4}$\cite{Nakamura:2010zzi}, we obtain
the corresponding $\delta {th}\equiv {th}-(M_D+M_{D^*})^2$ in
Tables~\ref{tab:2} and ~\ref{tab:3}. One can see that the frequently
used natural cutoff $\delta {th}=0$ is not supported by our
calculation. Notice that we distinguish the charged and neutral
$D(D^*)$ masses in our numerical calculation. To estimate the
uncertainty from $J/\psi D D^*$ form factor, we choose our power
parametrization and  the Gaussian form in
Ref.~\cite{RodriguesdaSilva:2003hh} for comparison. The
$s_1$-dependence of $g_{\psi''DD}$ is also taken into account by
adding the suppression factor $\exp(-I|\vec{p}_3|^2)$ into the
dispersive integral, where $I=0.4 \ \mathrm{GeV}^{-2}$ is extracted
from the charmonium mass shift~\cite{Pennington:2007xr}. It should
be stressed that the dispersive integral with the suppression factor
is not considered priority than the original one with a lower
effective threshold from the phenomenological viewpoint. Moreover,
both imaginary parts numerically decrease faster than $1/s_1$, so
the unsubtracted dispersion relation used here is self-contained.

In most of the parameter space, the dispersive part of the branching
ratio is dominant over the absorptive part for $J/\psi\eta$, while
for $J/\psi\pi^0$ both absorptive and dispersive parts are
important. The difference between different $J/\psi D D^*$ form
factor is small, despite that the absorptive part of our power fit
is systematically larger than the Gaussian fit in
Ref.~\cite{RodriguesdaSilva:2003hh} as expected. As shown in
Table~\ref{tab:2}, the branching ratios of $J/\psi\pi^0$ obtained
from the corresponding effective thresholds are in good agreement
with the experimental upper limit $2.8\times
10^{-4}$~\cite{Nakamura:2010zzi}. As an interesting investigation,
we take the threshold asymptotic to infinity with the suppression
factor, and the corresponding branching ratios of $J/\psi\pi^0$  are
still below the upper limit except for $g_{\psi''DD}^{low}$. In
contrast, the asymptotic limits of $J/\psi\eta$ are far beyond the
upper limit of the experiment $13\times10^{-4}$. Then, even taking
the suppression factor into account, the spectral information from
other resonances and continuum are still ambiguous, so that the
asymptotic limit is questionable and the effective threshold is
still necessary.

It is essential to recognize that the isospin symmetry breaking with
the vertex couplings is also an important dynamic source apart from
effects caused by the mass differences between the charged and
neutral $D^{*}$ mesons. In our formulation, the coupling
$g_{\psi''DD}$ and form factor $F_{D^*D\pi^0}(p_5^2)$ are extracted
from experimental data which suggest different values for the
charged and neutral couplings, respectively. Since the form factor
$F_{D^*D\pi^0}(p_5^2)$ can be better fixed by the experimental data,
the results listed in Tables~\ref{tab:1},~\ref{tab:2} and
\ref{tab:3} also reflect the effects from the isospin breakings of
$g_{\psi''DD}$. It is interesting to note that the larger absorptive
contributions actually favor smaller difference between the charged
and neutral $g_{\psi''DD}$ couplings, which is also observed in
Refs.~\cite{Achasov:1994vh, Achasov:2005qb} considering the
theoretical Coulomb correction for $g_{\psi''DD}$. Taking into
account the dispersive part, the central values of $g_{\psi''DD}$
give relatively small branching ratios for $\psi''\to J/\psi+\pi^0$,
while deviations from the central values can produce larger
branching ratios for $\psi''\to J/\psi+\pi^0$. Within the present
experimental uncertainty bounds~\cite{Nakamura:2010zzi}, the
predicted branching ratios for $\psi''\to J/\psi+\pi^0$ are at the
order of $10^{-5}\sim 10^{-4}$. Confirmation of this decay branching
ratio would be a strong evidence for the open charm threshold
effects in $\psi''\to J/\psi+\pi^0$. Note that our prediction of the
absorptive part is also close to the prediction of Ref.~\cite{
Achasov:2005qb} with isospin $I=0$ for $\psi(3770)$, i.e.
$Br_\eta(\mathrm{Abs})=8\times10^{-5}$ and
$Br_\pi(\mathrm{Abs})=2\times10^{-5}$, which is a consequence of the
similar values of the form factors in both approaches.

\begin{table}
\caption{Different $g_{\psi''DD}$ from experimental
data~\cite{Nakamura:2010zzi}.}\label{tab:1}
\begin{tabular}{|c|c|c|c|c|c|} \hline\hline
    &Br ($\psi''\rightarrow D^0+\bar{D}^0$) &Br ($\psi''\rightarrow D^++D^-$)&$g_{\psi''D^0\bar{D}^0}$&$g_{\psi''D^+D^-}$& Coupling Fraction (N/C) \\
    \hline
 $g_{\psi''DD}^{low}$&0.47&0.45& 12.26&14.62&0.84  \\
 \hline
  $g_{\psi''DD}^{cen}$&0.52&0.41&12.90& 13.95&0.92\\
  \hline
   $g_{\psi''DD}^{up}$&0.57&0.37&13.50&13.25&1.02 \\
\hline\hline
\end{tabular}
\end{table}

\begin{table}
\caption{The branching ratio without suppression factor.
``Abs'' denotes the absorptive part, and ``Tot'' is for the sum of
the absorptive and dispersive part. The flavor mixing angle is
$\alpha_P=38^\circ$. }\label{tab:2}
\begin{tabular}{|c|c|c|c||c|c|c|}\hline\hline
&\multicolumn{3}{c||}{ Our power fit }&\multicolumn{3}{c|}{ Gaussian fit ~\cite{RodriguesdaSilva:2003hh} }\\
 \cline{2-7}
& $g_{\psi''DD}^{low}$  &$g_{\psi''DD}^{cen}$ &$g_{\psi''DD}^{up}$  & $g_{\psi''DD}^{low}$  &$g_{\psi''DD}^{cen}$ &$g_{\psi''DD}^{up}$ \\
 \hline
 $Br_\eta$(Exp)($\times 10^{-4}$)&(5,\,13)&(5,\,13)&(5,\,13)&(5,\,13)&(5,\,13)&(5,\,13)\\
    \hline
 $Br_\eta$(Abs)($\times 10^{-4}$)&2.13&2.16&2.18&1.12&1.14&1.15\\
\hline
 $\delta th$($\mathrm{GeV}^2$) &(0.17,\,1.25)&(0.17,\,1.25)&(0.17,\,1.25)&(0.9,\,2.5)&(0.9,\,2.5)&(0.9,\,2.5)\\
 \hline
  $Br_\pi$(Abs) ($\times 10^{-4}$)&0.145&0.320&0.558&0.074&0.164&0.287\\
  \hline
  $Br_\pi$(Tot)($\times 10^{-4}$)&(0.46,\,1.36)&(0.33,\,0.35)&(0.71,\,1.13)&(0.49,\,1.60)&(0.17,\,0.23)&(0.50,\,0.81)\\
\hline\hline
\end{tabular}
\end{table}

\begin{table}
\caption{The branching ratio with suppression factor. ``Tot'' is for
the sum of the absorptive and dispersive part, and ``Asym'' means we
take the asymptotic limit $\delta th\rightarrow\infty$. Flavor
mixing angle is $38^\circ$.}\label{tab:3}
\begin{tabular}{|c|c|c|c||c|c|c|} \hline\hline
    & \multicolumn{3}{|c||}{ Our power fit } &\multicolumn{3}{|c|}{ Gaussian fit ~\cite{RodriguesdaSilva:2003hh} }\\
 \cline{2-7}& $g_{\psi''DD}^{low}$  &$g_{\psi''DD}^{cen}$ &$g_{\psi''DD}^{up}$  & $g_{\psi''DD}^{low}$  &$g_{\psi''DD}^{cen}$ &$g_{\psi''DD}^{up}$ \\
 \hline
 $Br_\eta$(Exp)($\times 10^{-4}$)&(5,\,13)&(5,\,13)&(5,\,13)&(5,\,13)&(5,\,13)&(5,\,13)\\
    \hline
 $\delta th$($\mathrm{GeV}^2$) &(0.26,\,1.60)&(0.26,\,1.60)&(0.26,\,1.60)&(1.15,\,3.50)&(1.15,\,3.50)&(1.15,\,3.50)\\
\hline
  $Br_\eta$(Asym) ($\times 10^{-4}$)&84.7&84.7&84.2&35.2&35.2&35.1\\
  \hline
  $Br_\pi$(Abs) ($\times 10^{-4}$)&0.145&0.320&0.558&0.074&0.164&0.287\\
 \hline
   $Br_\pi$(Tot) ($\times 10^{-4}$)&(0.48,\,1.49)&(0.33,\,0.37)&(0.69,\,1.05)&(0.53,\,1.73)&(0.18,\,0.26)&(0.47,\,0.73)\\
   \hline
  $Br_\pi$(Asym) ($\times 10^{-4}$)&13.76&1.47&2.86&5.58&0.62&1.25\\
\hline\hline
\end{tabular}
\end{table}

\section{Conclusion}\label{sec:4}
The ELA is very useful to investigate the nature of the near
threshold charmonia and charmoniumlike resonances. The largest
uncertainty of the ELA comes from the determination of the off-shell
effect, i.e. the form factors. In this paper, we investigate the
isospin violating decay of $\psi(3770)\to J/\psi+\pi^0$. In this
process, there is only one $D$-meson loop contributing to the
absorptive part, and the form factors encountered in the loop
calculation can be relatively well controlled. With the help of
QSSR, we extract the $J/\psi DD^*$ form factor as an implement from
the first principle of QCD. The $DD^*\pi^0$ form factor can be well
determined from the experimental data of $D\rightarrow\pi l\nu$,
which has been measured with high accuracy. We also explore the
dispersion relation to evaluate the dispersive part of $\psi(3770)$
non-$D\bar{D}$ decays, and find they take an important part in most
of the parameter space. It means that the IML effects below the open
charmed meson threshold cannot be neglected in general. Different
from the traditional natural cutoff of the effective threshold in
the dispersive integral, we extract them from the experimental data
of $\psi(3770)\rightarrow J/\psi+\eta$. Our prediction of the
branching ratio of $\psi(3770)\to J/\psi+\pi^0$ is less than
$3\times 10^{-5}$ with the couplings $g_{\psi''DD}$ extracted from
the central values of the data. Within the experimental uncertainty
bounds for the extracted $g_{\psi''D^0\bar{D^0}}$ and
$g_{\psi''D^+D^-}$, the branching ratio of $\psi(3770)\to
J/\psi+\pi^0$ can reach the order of $10^{-4}$. Notice that the
understanding of the isospin violation of $g_{\psi''D\bar{D}}$ is
not a trivial task despite the Coulomb correction favors the
experimental central value~\cite{Voloshin:2007dx}. It is also
suggested that a small admixture of isovector four-quark component
of $\psi(3770)$ may also give a measurable decay rate to
$J/\psi\pi^0$~\cite{Voloshin:2005sd}. In Ref.~\cite{Rosner:2004wy}
the four-quark component is viewed as a reannihilation effect of
$D\bar{D}$. To some extent, the nature of $\psi(3770)$ hides in the
coupling $g_{\psi''DD}$. Meanwhile, the forthcoming BESIII
measurement of $\psi(3770)\to J/\psi+\pi^0$ will be able to provide
useful information about the QCD motivated form factors and clarify
the role played by the IML. We plan to discuss the isospin
violations with the charged and neutral couplings $g_{\psi''D^+D^-}$
and $g_{\psi''D^0\bar{D^0}}$ elsewhere.

\acknowledgments
This work is supported, in part, by the
France--China Particle Physics Laboratory, National Natural Science
Foundation of China (Grant No. 11035006), Chinese Academy of
Sciences (KJCX2-EW-N01), and Ministry of Science and Technology of
China (2009CB825200).

\end{document}